\documentclass[sigconf]{acmart}
\AtBeginDocument{%
  \providecommand\BibTeX{{%
    \normalfont B\kern-0.5em{\scshape i\kern-0.25em b}\kern-0.8em\TeX}}}

\setcopyright{none}
\renewcommand\footnotetextcopyrightpermission[1]{}
\usepackage{multirow}
\usepackage{array}

\usepackage{balance}       
\usepackage{graphics}      
\usepackage[T1]{fontenc}   
\usepackage{mathptmx}
\usepackage{color}
\usepackage{booktabs}
\usepackage{textcomp}
\usepackage{makecell}
\usepackage{xspace}
\usepackage{fancyvrb}
\usepackage{amsthm}
\usepackage{hyperref}

\usepackage{microtype}        
\usepackage{ccicons}          

\begin{document}

\title{Continuous Pupillography: A Case for Visual Health Ecosystem}

\author{Usama Younus}
\authornote{Scholarly article by Usama Younus, supervised by Prof. Nirupam Roy}
\email{uyounus@umd.edu}
\affiliation{%
  \institution{University of Maryland}
  \city{College Park}
  \state{Maryland}
  \country{USA}
  \postcode{20740}
}

\author{Nirupam Roy}
\email{niruroy@umd.edu}
\affiliation{%
  \institution{University of Maryland}
  \city{College Park}
  \state{Maryland}
  \country{USA}}

\renewcommand{\shortauthors}{Younus and Roy}
\begin{abstract}
This article aims to cover pupillography, and its potential use in a number of ophthalmological diagnostic applications in biomedical space. With the ever-increasing incorporation of technology within our daily lives and an ever-growing active research into smart devices and technologies, we try to make a case for a health ecosystem that revolves around continuous eye monitoring. We tend to summarize the design constraints \& requirements for an IoT based continuous pupil detection system, with an attempt at developing a pipeline for wearable pupillographic device, while comparing two compact mini-camera modules currently available in the market. We use a light algorithm that can be directly adopted to current micro-controllers, and share our results for different lighting conditions, and scenarios. Lastly, we present our findings, along with an analysis on the challenges faced and a way ahead towards successfully building this ecosystem.
\end{abstract}


\keywords{Internet of Things, Pupillography, Ophthalmology, Wearable Devices, Remote Analytics, Medical Diagnostics}


\maketitle
\pagestyle{plain}

\section{Introduction}

Vision, being one of the core senses of a human body, plays a pivotal role in our daily life. It is said that about 80\% of our total perception of the world is through our vision \cite{gregory1998brainy}. To put it in another perspective, the mastery of our hand cannot supersede beyond what the vision perceives; and the bodily organ that assists in this process of worldly perception is our eyes. Through them, we not only comprehend most of our surroundings, but also ensure a healthy sustainability of our entire being. In-case of failure of other senses such as hearing and smell, it is our vision that protects us from vision. However, the eyes are not only a means to explore the world that is external to our body. This organ can also aid in analyzing and visualizing our internal self i.e. our personal health conditions. 

The eyes are not the only doorways to our internal health. There is an extensive research into using other senses, and bodily parts to delve into our internal realm. A great number of various applications and diagnostic devices have been developed utilizing the different functionalities of our body. Many of these applications are supported by the accelerated growth of the IoT space, and a continuous trend towards miniaturization of sensors and processing tools. The IoT device space itself is experiencing an exponential growth with an estimated growth size of \textbf{USD 1463.19 billion} in 2027, which follows a compound annual growth rate (CAGR) of \textbf{24.9\%} \cite{(IoT)MarketSize}.

The biomedical diagnostic and therapeutic device industry is not far behind in its adoption rate and market size. This industry itself is expected to have a market cap of around \textbf{USD 111.9 billion} by 2028, which shows a CAGR of \textbf{26.8\%} from 2021-2028 \cite{WearableDevicesMarketSize}. This signifies a shareable size of \textbf{7\%} with respect to the IoT industry. A number of smart devices are already taking the market by storm and have found audience-wide acceptance and integration into our lives. A few of these devices are summarized in the table \ref{tab:Wearable_examples}.

\begin{table}
\centering
\begin{tabular}{>{\centering}p{0.46\linewidth}>{\centering\arraybackslash}p{0.46\linewidth}}
\multicolumn{2}{c}{\textbf{Applications of Wearable IoT Devices}}\\
\hline
\begin{itemize}
  \item Wearable Fitness Trackers
  \item Smart Health Watches
  \item Smart ECG Monitors
  \item Wearable Pulse Oximeters
  \item Smart Thermostats 
  \item Wearable Blood Pressure Monitors
  \item Smart Shoe Sensors
  \end{itemize}
  &
\begin{itemize}
  \item Wearable EEG Monitors
  \item Wearable Eye Tracking
  \item Wearable EKG Monitor
  \item Biosensors 
  \item Biomedical Rings
  \item Smart Hearing Aids
  \item Smart Sleep Tracker
  \item Wrist Actigraphs
  \item Headbands
  \end{itemize} \\
\hline
\end{tabular}
  \caption{Smart Wearable Health Monitoring Devices. \cite{WearableDevicesMarketSize}}
  \Description{A number of IoT examples in the wearable biomedical space}
  \label{tab:Wearable_examples}
\end{table}

Within this space, vision and eyes hold a significant importance, as they account for 80\% of our perception \cite{gregory1998brainy} stated earlier. To fully leverage the user experience of wider applications through IoT, it is imperative to fully leverage the complete functionality of the eyes for future IoT devices and applications. However, not many applications have attempted to bridge this gap of wearable technology and visual applications. Some groups have researched on the application of Eye-ball tracking for Virtual Reality headsets \cite{clay2019eye}, Gaze-detection for identification of Glaucoma \cite{li2019attention}, Retinal images for Blood Sugar measurements \cite{EyeTestforDiabetes}, Person Identification through Iris Recognition \cite{bowyer2016handbook} etc.; but miss out on an important aspect of the eye i.e. Pupils, which can contribute to a good number of ophthalmological diagnostic applications (which are discussed in next sections). 

In this paper, we address some of the possible applications that can be funded through the use of continuous pupilliography. We discuss some of the prior work and open challenges in the space of wearable Pupillography. We also make an attempt with minimalistic budget and off-the-shelf resources to create a pipeline that can establish a framework for designing a holistic system of measurement and post-processing of pupils. The algorithm used is a light-weight implementation of Hough transform, which gives significant results given the constraints. This can allow the users to completely manage and timely respond to their varying health conditions through the scope of their eyes.

\section{Biology 101}
In this section, we tend to briefly cover the components of an eye, specifically pupils and the connected nerve pathway that triggers the response of pupils in different visual scenarios. This pathway, which interconnects both of the eyes with perception cells in brain, is a gateway to visual health and can be utilized for such diagnostic purposes.

\subsection{Understanding Eye Pupil: Overall Structure \& Functionality}
\begin{figure}[h]
  \centering
  \includegraphics[width=\linewidth]{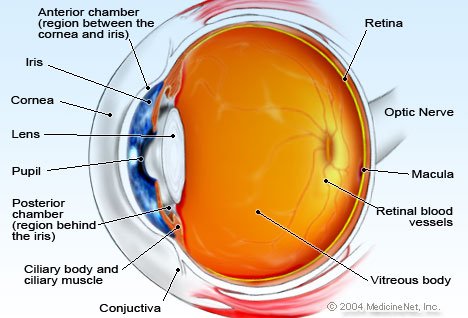}
  \caption{A labelled cross-sectional view of the complete eye structure. Exactly reproduced from the Medical Illustrations of \cite{EyeAnatomy}}
  \label{fig:eye_figure}
\end{figure}

The eye, which is the forefront of our visual perception system, consists of a number of sub-parts. All of these parts can be clearly seen in the figure \ref{fig:eye_figure}. These components can be described as:
\begin{itemize}
    \item \textbf{Cornea:} This is the transparent section of the eye which is involved in processing and focusing of light. It lies just in-front of the eye, and is often invisible to general view. Wearable lens rest on this part of the eye.
    \item \textbf{Iris:} This is the circular colored section of the eye, which helps in the regulation of the amount of light passing through eyes. This also forms a unique imprint for each person
    \item \textbf{Pupil:} This is the dark central spot within Iris, which controls the light passage and amount. It dilates and contracts depending on the amount of brightness in the environment. 
    \item \textbf{Retina:} This is a very thin lining of nerve tissue layer at the back-end of the eye. It is simulated through the reception of light and sends impulses to the virtual cortex of the brain through optic nerve
    \item \textbf{Macula:} This is a small central area within retina which contains specialized photo-receptive cells that aid in distinguishing very minute details.
    \item \textbf{Optic Never:} This nerve connects the eyes with the brain, and carry impulses from retina to visual cortex.
    \item \textbf{Vitreous:} This is the jelly-like material which comprises the center of the eye. It is a colorless.
\end{itemize}

Out of all these elements, the most concerned one for us in this article is the pupil, which servers to control the inlet of light into the eyes. The pupil functions in a passive response manner, where the aperture of the pupil correlates with the amount of light falling onto it. In a darker setting, the pupils of a person are larger in diameter to capture the maximum amount of light in the environment, perhaps around 8 mm or more. However, as soon as the room is lighted up, there is an immediate response by the pupils in the form of quick contraction, which is called the light reflex. This reflex of the pupils is bilateral, which means that even if only one of the eyes out of both is exposed to light, the rate of contraction for the both the pupils is nearly similar. This aspect of pupillary response is of significant importance and is the most relevant reason for using pupils in subsequent daignostic applications. 

After certain time, both the pupils expand even in the presence of the light, with the final size proportionate to the quantity of light illumination - with a resting state of around 4mm to 5mm. The pupillary contraction can also occur when a person tries to view a very close-by object, which is termed as near reflex. In bright light, the pupils serve for the additional purpose of focusing the depth of the optical system by reducing the aperture. and controlling the aberrations (failure to capture and focus rays onto the retina. This functionality is however negligible in dark ambiance, where aberrations need not to be accounted for and only maximum capture of light is needed. 

The dilation of pupil can also occur as a result of strong emotional stimulus or whenever a sensory nerve is simulated; this resulting in dilation responses in the case of pain or fear. This dilation and contraction of the pupil is brought about by the muscles of the iris. Additionally, the dilatory muscle of the iris is regulated by the nerve fibers of the sympathetic system - any simulation of these nerves in neck can also bring about a significant dilatory response in pupil. The sensory pathway for general light reflex of pupil involves cones, rods, biploar cells, and ganglion cells. 


\subsection{Pupillography: Current Measurement Methods} 
Having covered some of the structural details of the eye and pupil reflex mechanism, we move onto the process of measuring the reflexive action of pupils, called \textbf{Pupillography}. There are a number of different ways through which the response of pupils is estimated and its pathological implications are inferred. The techniques differ in their subjectivity and objectivity, ranging from flash-light test which is a completely subjective methodology based on the clinician's practice and experience, to infrared video pupillography which automates the measurement process by involving extensive hardware. A brief overview of these is given below:
\begin{itemize}
    \item \textbf{Swinging flash-light test:} This testing mechanism uses an indirect opthalmoscope or a flashlight to capture the response. The aim is to keep the illumination in range such that the pupil responds linearly the light simulation, with the general practice being varying the distance from the eyes. The examiner needs to ensure that the light makes a 45 degree angle with the optical axis to restrict sudden reflex moments (thus missing out key defects). The linear correlation stops to exist for diameter sizes of less than 3-4 mm \cite{wilhelm2011disorders}.
    \item \textbf{Infrared Video Pupillography:} This technique uses an automated measurement of the pupillary response through illumnation of light by an infrared source and capturing that response through video cameras. An afferent pupillary defect (APD) is measured after exposing both eyes to varying stimuli, in order to bring about a same amplitude. This method correlates the amplitude with the light reflex, however involves very costly and extensive equipment. It was first introduced in  \cite{eyekinetix}
    \item \textbf{Electrooculgraphy:} This technique measures the voltage potential difference between the retina of the eye and the cornea. In this method, electrodes are placed closer to the eye at fixed points, and the varying electric field is measured resulting from the movement of the eye. Though this test is not a direct measure of pupillary response, however certain applications use pupillary reflection as an intermediary step in the process \cite{macneil2020tracking}
\end{itemize}

There are a number of other clinical methods used for the diagnostic purposes of pupillary response. In general, most of these techniques are comparable (to some extent) in their efficiency, cost, timing, and required domain expertise, however none of them are designed for prolonged measurement periods; thus leaving opportunity for designing a compact system in this space of ophthalmology.
\section{Potential Applications \& Prior Work}
Through our earlier illustration of the exponential growth of IoT and Biomedical Wearable space, we can fairly make an assumption that a significant amount of effort and resources would be deployed for building holistic solutions around visual ecosystem in the near future. In this section, we explore some of the applications that can be favoured by such a system and some of the prior works that are limited in their design considerations, scope, and application.

\subsection{Pupillography Use-Cases}
As discussed earlier in the Biology 101 section, the process of pupillary response can expose a number of conditions not only related to the light reflex. The bilateral response of the pupils due to common nerve endings can identify a number of additional ophthalmological and psychological diseases, not easily observable through a number of other diagnostic tests. This process can serve as a preliminary/early detection mechanism, which can support extensive subsequent testing procedures.  Though there is an ever-growing list of applications that can benefit from a continuous system of visual response measurement, we have mentioned a number of them here to highlight and signify its importance.

There has been a significant amount of research that correlates the pupillary response with early detection of a number of diseases \cite{wilhelm2003clinical}. Some research has linked the decrease in the pupillary response rate with the effects of toxicology i.e. \textit{ocular toxicology} \cite{wilhelm2003clinical}. This includes the effects of antidepressants, anti-hypertensives or amphetamines on the ocular behaviour, and is also an active field of research in drug effects measurement. Another application which has recently received some attention in IoT space is evaluating the quality of sleep. Research by Mier H. ~et al. \cite{kryger2010principles} shows that irregular response rate of pupils can be one of the bio-markers for detecting these patterns. \textit{Glaucoma} is a condition in which the optic nerve is damaged, thus leading to poor vision.  It is one of the leading causes of blindness in people over 60, and is generally tested through both the flash-light test and infrared video test, using APDs as biomarkers. One article by Bisant A. ~et al. in the magazine of \textit{Review of Optometry} shows that an early correlation can be established with APD testing, thus pupillography \cite{DoAPDsMatter}.

\textit{Narcolepsy}, a chronic disorder of sleep, also finds its early symptoms in the varied reflexive response of the pupils, and slight consistent deviation in the aperture size. People with this order tend to find overwhelming drowsiness during the daytime and experience sudden episodes of sleep attack \cite{daroff2014encyclopedia}.\textit{Anisocoria}, which refers to the condition of one eye differing from the other is size, is also detectable through clinical examination of the cranial nerves (reflected through the pupillary response) \cite{tubbs2015nerves}.\textit{Holmes-Adie syndrome} is a rare syndrome that leads to larger pupil diameter than normal (dilated), and slower reaction times to stimulation by direct light \cite{AdieSyndrome}.

\textit{Parinaud syndrome}, a brain stem disease, effects the ability of the eye to move up and down, which can have pupillary hyporeflexia (less responsiveness to stimulus) as its clinical deterministic symptoms \cite{lai2010brain}. \textit{Horner syndrome} can show multiple symptoms signifying its clinical origin. This syndrome is the result of half-sided disruption of nerve pathways leading from the brain to the face in a body, and can be partially characterized by the smaller size of the pupils \cite{eggenberger2014neuro}. Lastly, a good number of recent publications highlight changes in the pupillary response rate as one of the bio-markers for early risk of \textit{Alzheimer disease} \cite{TwoviewsonAlzheimersbiomarkers,WhatAlzheimersHastoSayaboutEyesight,risacher2020visual,granholm2017pupillary, kremen2019pupillary} - which in itself is a significant discovery. Being the sixth leading cause of death in USA, this disease is characterized by an irreversible decline in the cognitive abilities, which further leads to memory loss and inability to perform daily activities.

\begin{figure}[h]
  \centering
  \includegraphics[width=\linewidth]{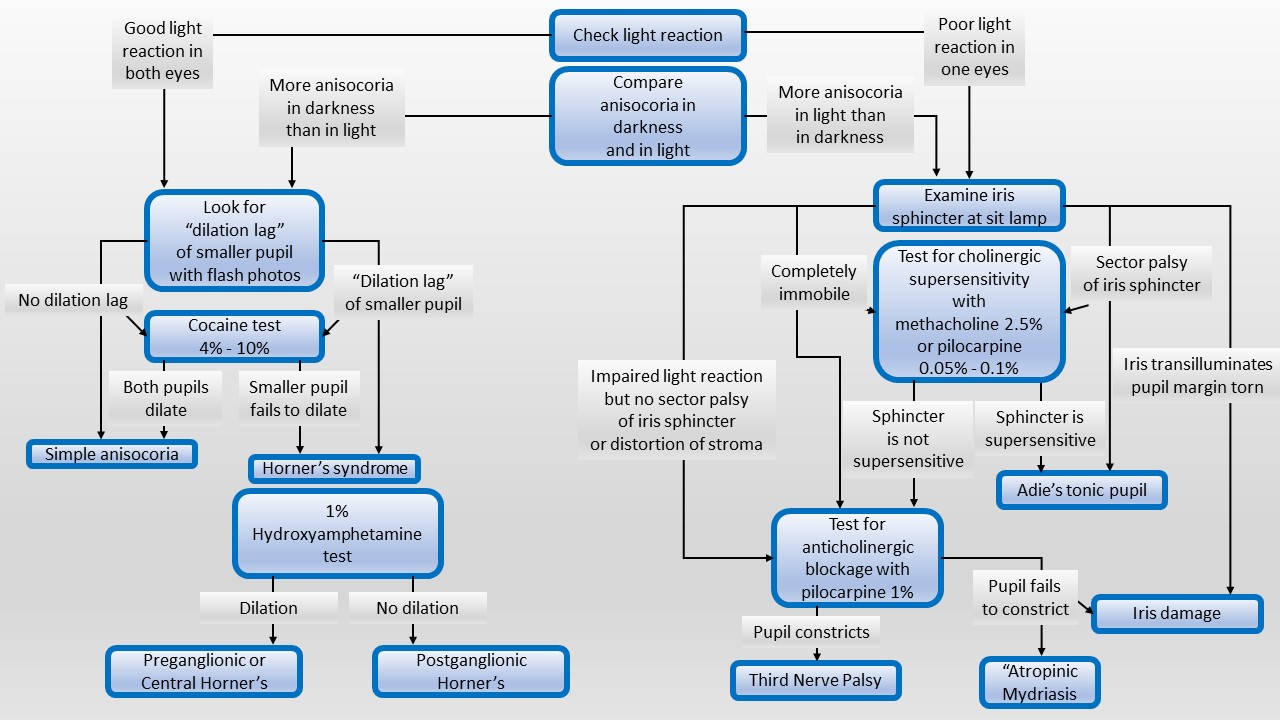}
  \caption{A Systematic Flow Diagram of Pupillary Response Rate and Subsequent Categorization of Various Diseases \cite{palmer2007assessment}. }
  \label{fig:disease_chart}
\end{figure}

\subsection{Some Prior Work}

Using IoT based solutions for visual diagnostics is not a completely new implementation. A number of prior works have explored slightly different variations and applications, in contrast to the one we are proposing in this paper. Alex Mariakakis ~et al. \cite{mariakakis2017pupilscreen} suggest a smart phone based mechanism that can be used to measure the pupillary reflex rate for assessing traumatic brain injury. Their proposed design uses a 3D printed module that serves as a holder for the phone. The images are then taken through phone camera and processed on the same device. Though their approach has high processing capabilities and camera resolution, it cannot be easily translated to perform continuous monitoring due to the head-mounted camera that blocks the view. Kurtis Sluss ~et al. \cite{sluss2019evaluation} propose a hand-held device to measure the same phenomenon. This device has slightly better analytical prowess than PupilScreen \cite{mariakakis2017pupilscreen}, however the device itself is a customized biomedical gadget comparable to the size of a hairdryer, and is costly to reproduce. Additionally a similar limitation for adapting it for continuous monitoring reduces the design option for a visual health ecosystem. 

Similar limitations exist in the works of Taehying Kim ~et al \cite{kim2020experimental} and Mei-Lan ~et al \cite{ko2014design}; both of which attempt at utilizing pupillary response for visual fatigue analysis and neuropathy of diabetic patients respectively, but do not have special design considerations for their devices themselves. An interesting design approach is deployed by Joon Hoon ~et al. in their paper \cite{lee20203d} where they attempt to develop a multi-functional e-glass that can measure a number of signals, including EEG, EOG, UV etc. We believe that their design can be modified and enhanced further to include a pupillary measurement module. The device currently lacks any cameras for direct evaluation of the light relfex of pupils. 

Lily Yu-Li ~et al. \cite{chang2017infrared} use an off-the-shelf pupillometer with a combination of a LED probe (powered by a smart phone) to simulate the pupils and record a relational response. Through their reported accuracy is high for resting pupil size, however, the solution lacks portability capabilities, and their design approach is unusable for a consistent monitoring system. Lastly, Ou-Yang ~et al.'s solution \cite{mang2016wearable} tries to strike a good balance between portability, and computational efficiency. Their developed prototype has two separate systems: a camera based system for capturing images for analysis, and a lighting system that provides LED lights of varying wavelengths (including IR) for simulation. The only open question to their implementation is the requirement of a secondary lighting source to the natural environment to measure the pupil size.

Though all of these proposed solutions tackle different aspects of pupillography and tend to use them for a varying number of problem statements, most of them are unsuccessful in defining or deploying the design requirements for a ubiquitous monitoring system. We tend to briefly explain these design constraints in our next section to show they restrict the possible solution space for this system.

\section{Experimentation}
In this section, we aim to gain a deeper understanding of the design requirements for developing ubiquitous systems. These requirements could improve the existing efforts to be more complete, realistic, and user centered. To achieve this goal, we explore it through design thinking i.e. analyzing the design constraints for an inclusive IoT based solution, and different design options. After shortlisting some of the options in the market that fall within these design parameters, we proceed to perform a number of experiments covering a wide range of scenarios and conditions.

\subsection{Design Constraints}
Starting with the vision of ubiquitous computing \cite{weiser1991computing}, the IoT space has evolved on the technological progress of bringing increased miniaturization and availability of information and communication technology at decreased cost and energy-consumption. In order to understand the mechanism of the design, we need to first look into the whole pipeline of the processing system, and evaluate the respective constraints of each part.

The system has two basic functionalities: it needs to capture the picture of the eyes with a resolution clear enough for proper post-processing and it the next step is to post-process the image, find the region of pupil and compute the diameter. Based on this simple but coherent pipeline, we now discuss a number of design decisions that need to be considered:

\begin{enumerate}
    \item \textbf{Compactness} The camera module should be compact enough that it can be easily adapted for a remote or continuous setting. This ensures that the design is light-weight enough for a wearable device application and helps in regulating the cost aspects as well. 
    However, this decision is deeply tied with the resolution capacity as mini-camera modules lack the feature of auto-focus and thus can lead to poor quality images not usable in post-processing. We will see such results in one of our experiments. 
    \item \textbf{Illumination Source} The camera module needs to operate in two different environments: within a sufficiently lighted space and within very low to none lighting. As we will see in our findings section, a need for an LED light source is present for very dim lighting environments. 
    \item \textbf{Algorithmic Complexity} The post-processing technique should use as low computational resource as possible. This is due to the fact most standalone micro-processing units are limited in their computational capacity. For our experiments, we have used a very light implementation with bare-minimum requirements. 
    \item \textbf{Power Requirements} This design parameter is deeply tied with the compactness a camera and algorithmic capability of the micro-processing unit. The higher these two requirements are, the higher the need for more electrical power. 
    \item \textbf{Direct/Indirect Measurement: Approach} Lastly, this is a design choice that we touch-based to give another perspective to the design problem.  The design constraint deals with the approach that we would like to adopt for capturing the images: either a direct mechanism with cameras directed towards the eyes, or a indirect method of capturing the image of pupils through any reflective surface, such as glasses. We discuss this option as well in our hardware selection section. 
\end{enumerate}

We have put forth a number of design constraints that need to be addressed for proposing a completely wearable solution. These constraints can be used as the guiding principles to narrow down the design options and the possible solution space to explore.

\subsection{Proposed Solution}
Here we put forth the components of our initial prototype. We work through a number of available options present in the market, and use our previously defined design constraints to narrow down our selection process. We end up with experimenting with two camera modules because of the poor image quality of one in day-light. We also try different lighting sources to establish the case for requiring an external light source for illumination purposes. For our processing algorithm, we target low computational complexity over other options, as this would ensure we can run our technique on a micro-controller.

\subsubsection{Hardware Selection} The hardware components of our project compromise of the camera module, the illumination source, processing unit, and power source. We discuss each of these options separately below:

\begin{figure}[h]
  \centering
  \includegraphics[width=\linewidth]{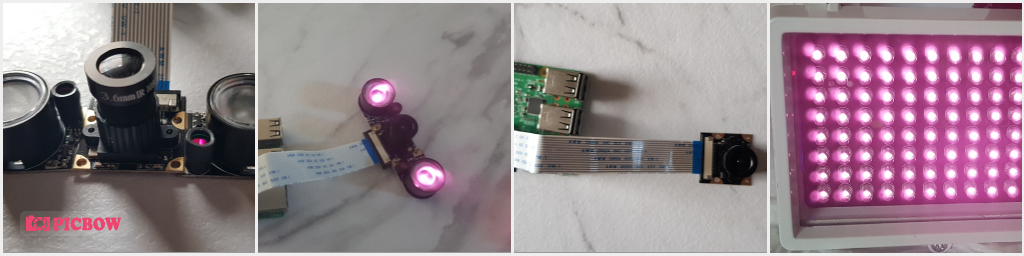}
  \caption{From Left to Right: 5MP ArduCam, ArduCam with IR Illumination, 7MP Pi Noir V2, LED Illumination Panel}
  \label{fig:equipment_chart}
\end{figure}

\begin{enumerate}
    \item \texttt{Raspberry Pi 3:} For the processing unit, we had the option of choosing between Adafruit PyBadge or Raspberry Pi 3. Though the former comes with ML capabilities, we went with the later option which is a slightly older model in the raspberry micro-controller line. Pi 3 has a compact form in comparison to PyBadge and the per-unit cost is also around \$30. It is easily programmable for different functionalities, and has around 1.2GHz processing power and 1GB of RAM. Another aspect to it is the comparable processing capacity to other micro-controllers, such as PIC. This means any development on this can be easily extended to other controllers. 
    \item \texttt{5-Volts Supply:} We powered the micro-controller through a 5-volt USB-C cable tethered to a laptop. This can be replaced with a power adapted of similar voltage. This again gives us an option to have an entire battery powered setup in the later stages, which doesn't rely on any live power sources.
    \item \texttt{Direct Approach:} For our set of experiments, we selected the direct approach to capture the images. This approach also impacted our camera selection and algorithm development. For the indirect approach, the available option was to coat the inside of glasses with a IR or Photo-reflective film that can reflect the pupil image in the glass - from where we could capture the image. 
    Current items in the market for coating purposes included thermal insulation, IR blocking film, liquid nanotint solar coating, absorber coating etc. The only draw-back to these was that they hinder the sight of the person wearing the glasses, and can produce reverse flashing on the inner coating which impacts the all-day usage of this system. Another option was to use a ball-eye reflector to capture the image, however we dropped that option given the time and resouce contraints of this project.
    \item \texttt{Pi Noir V2 \& Arducam:} This constraint posed another challenge in our selection process. Given our initial selection of Raspberry Pi, we had to opt for an option that interfaced well with the design process. We had a number of options ranging from small pinhole camera to monochrome USB no-infrared camera modules. A few options revolved around mini-cube cameras, as well as using spy-cameras for their compactness in size. 
    We went ahead with two options of Pi Noir V2 and Arducam. The Pi Noir is an infrared insensitive module as its IR filter has been removed. It comes with 7MP of resolution, however doesn't have any auto-focus. To consider this drawback, we selected another camera module of Arducam, which is a 5MP camera (lesser resolution) but has an adjustable focus. As seen in experiments, this feature led to better quality pictures as a result.
    \item \texttt{Infrared Panel:} For considering different scenarios, and observing the performance of the cameras under a number of lighting conditions, we also used a 5-Volt Infrared LED panel consisting of an array of more than 100 IR LEDs. We ran a number of experiments under this light, and also used the in-built IR light source of Arducam (2 LEDs) in one set of experiments. 
\end{enumerate}

All of these components can be seen in the Figure \ref{fig:equipment_chart}.

\subsubsection{Processing Algorithm} The selection of the post-processing technique revolved around the requirements for keeping the computational complexities lower. This was also highly correlated with the selection of micro-controller, as PyBadge came with a pre-installed suite of libraries for such purposes. Since we selected Pi 3 as our preferred processing unit, our initial challenge was to use a technique that can be easily implemented on the board, and additionally can be easily transferred to other controllers with minimum loss of functionality. Henceforth, we chose to use a conventional transform technique over advanced machine learning methods due to their compute requirements. This technique, which is deployed for the use finding circles, is called \textit{Circle Hough's Transform}. This is a technique which extracts circles from imperfect images, while serving as a feature extractor. Since our experiment involved low quality and distorted images as well, this technique can be adopted to give good quality detection. 

\begin{equation}
    (x - a)^2 + (y - b)^2 = r^2
\end{equation}

This method uses the equation of circle with varying starting points in 2D space, with \textit{"a" \& "b"} being (x,y) coordinates and \textit{"r"} being the radius. This further implements an accumulator matrix and a voting mechanism. Initially, the code extracts a number of enclosed circles in the image space. It then draws a number of circles with fixed-radius on the surface of the enclosed surface. The next step involves searching for the intersection point these circles, which corresponds to the origin point of the original circular patch under consideration. This process is repeated over varying radius lengths for each enclosed surface, and the central points are stored in an accumulator matrix of same dimensionality (as the image). The algorithm then runs a voting mechanism to select the origin point with maximum number of votes (intersections). The entire process involves searching through 3D space, as it finds the three unknown parameters of \textit{(a, b, r)}.

Our algorithm works by taking in the images captured through our camera module, and post-processing it into a gray-scale image. Then we apply a \textit{Gaussian Blur} filter of kernel size 5x5 to reduce the noise inherent in low quality images, as well as to smooth edges and consider the voting process on each pixel. We then extract the median Blur from our image to cater for the \textbf{salt \& pepper} noise in the image still remaining after \textit{Gaussian Blurring}, and specify a certain threshold using \textit{Inversion Thresholding} (in range of 25 - 40 generally) for converting images into binary valued. This process is crucial as it drops most of the access circles and irregularly enclosed surfaces in the image, thus leaving behind usually the pupil and retina. We then apply the \textit{Hough's Transform} by building a \textit{Retrospective Tree (RT}. 

There are a number of ways to deal with concentric circles in Hough Transform, whereas each differ by the hierarchy of contours returned by them. RT returns all circles in a nested hierarchy, with each nested circle properly indexed with the outer one. Other options include obtaining only the outer-most circles in concentric circles, or obtaining a list of all circles without any hierarchical distribution. Since the images of eyes contain the outer contour of retina, as well as other enclosed shapes of retinal patterns, RT provides a good measure of obtaining these indexes. We then sort these indexes and select the one corresponding to the pupil. Lastly, the diameter of this contour is computed using the contour-area, and finally the original enclosed shape in plotted on the image, along with the center of the circle.

The computational prowess of this light-weight implementation originates from the fact that this only uses two libraries for the entire operation: \texttt{Python} and \texttt{OpenCV2}. The python library is used as a wrapper for our entire code base, while the OpenCV2 has a lite version available for Raspberry Pi 3 (which makes it a perfect scenario for our use-case)
\section{Findings \& Discussion}
In this section, we present the results from the number of experiments performed. We analyze the performance of our algorithm, as well as the capabilities of the hardware selected. We share the diameter values for each of the contours identified in those images. We then move onto discussing the results, and the drawbacks discovered in different settings to validate our design constraints put forth earlier.

Note: Individual images from different experimental setups were cropped and compiled into 9x9 image grids, using a free online collage tool referred to as "PicBow" \cite{Picbow}. Different images were taken at different lengths from eyes.

\subsection{Experimentation Results}
In total 6 sets of experiments were performed following different combination of the hardware available. We had previously employed a mixture of two people for the data sampling process, ensuring diversity of retinal colors \textit{black \& brown}. Since our initial proposal was to validate the case for a continuous monitoring system, we manually captured the pictures with varying distances to capture the relevant effects. We also performed experiments in natural light, room light, LED panel light, as well as the camera module IR light. To also add some diversity to the experiments, we tried to evaluate different eye positions and the efficiency of our system in finding the pupils. 

The algorithm encloses the surface identified, computes the intersection point (i.e. the origin of circle), along with the radius and plots a circle of that radius on the origin. The 6 set of experiments that were performed can be categorized into following settings:

\begin{enumerate}
    \item \textbf{PI Noir V2 (Person One):} These set of experiments were performed using the Pi Noir V2 camera under different lighting conditions, including daylight \& room light with our first person. The results can be seen in the Figure \ref{fig:results_old}. We can see from the images that the picture quality of the Noir camera is not very good under natural lighting sources. This is leading to white patches, right besides the pupil, thus rendering it difficult to find the pupils. Varying thresholds were applied here, ranging from 20 - 80, however the adjacent white patch enjoins the circle in computation. 
    The slight color variation is due to the IR filter being removed from this camera model.
    
\begin{figure}[h]
  \centering
  \includegraphics[width=\linewidth]{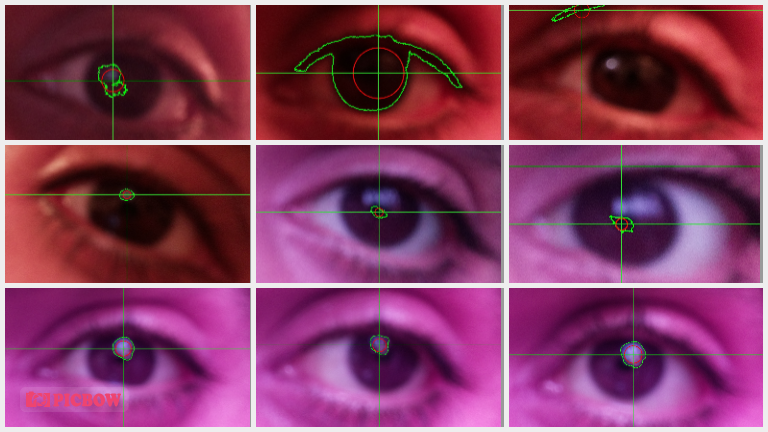}
  \caption{\textit{Noir V2 Person One:} Images captured with PI NOIR V2 camera. The first four images from top left to middle first were taken in room setting. The remaining images were captured in daylight}
  \label{fig:results_old}
\end{figure}

    \item \textbf{PI Noir V2 (Person Two):} These set of images were taken during the time of dawn in morning. The algorithm is slightly better able to distinguish the pupils from the white patches due to the lighting conditions. However, the results signify that Pi Noir camera wasn't able to completely capture crisp sharp images of the eyes. Additionally, since the camera doesn't have any focusing module, the images from a distance perform poorly than closer ones.
    
\begin{figure}[h]
  \centering
  \includegraphics[width=\linewidth]{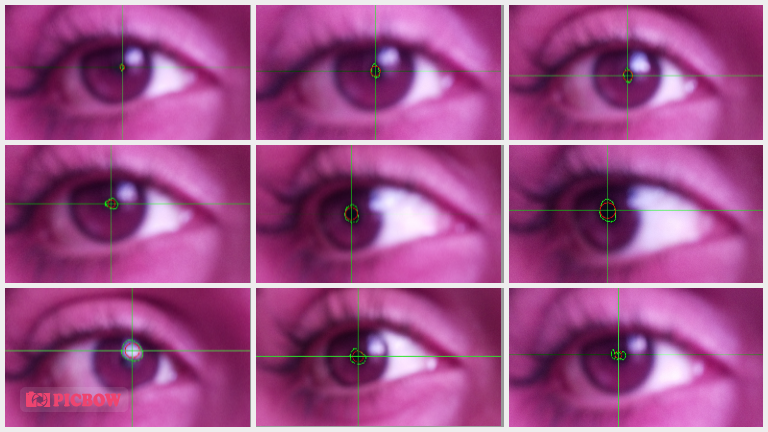}
  \caption{\textit{Noir V2 Person Two:} Images captured with PI NOIR V2 camera during early morning.}
  \label{fig:results_new}
\end{figure}

     \item \textbf{PI Noir V2 with IR Lamp:} For this set of experimentation, we deployed the 5-volt LED lamp as an illumination source in a completely dark room. We can clearly see a starch difference between the images taken in Figure \ref{fig:results_led_lamp} and all previous images. The image quality of the Pi Noir camera is immediately enhanced with the IR source. Even for different orientations of the eye, the algorithm is able to almost precisely capture the size of the pupil. This will be later seen in the diameter table as well, where the mechanism is able to capture the minute differences in the circles, and these diameters enhance as the pupils dilate in the images.
     
\begin{figure}[h]
  \centering
  \includegraphics[width=\linewidth]{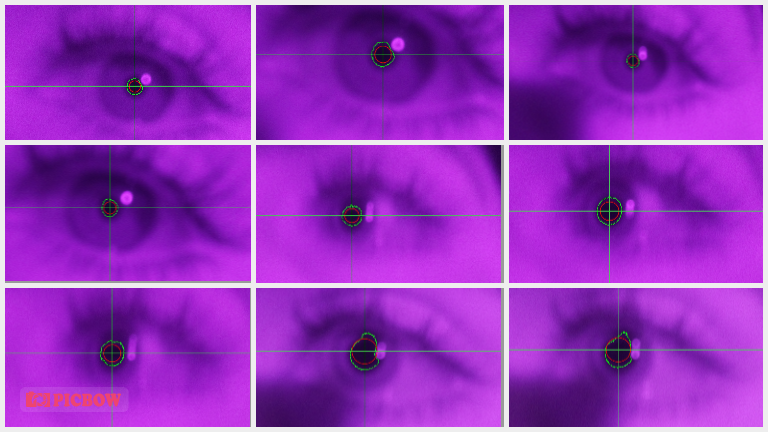}
  \caption{\textit{Noir V2 with IR Lamp:} Images captured with PI NOIR V2 camera under IR illumination source.}
  \label{fig:results_led_lamp}
\end{figure}

    \item \textbf{Arducam 5MP Camera:} The images in Figure \ref{fig:results_new_cam} were captured through the 5MP camera module developed by Arducam. We can assess from these images, that the color contrast of this camera is very different from the previous ones, with a preference towards the brighter shade. 
    Since this module comprises of an adjustable focus, we corrected the focal length for all the varying distances. This ensured very sharp images, thus leading to very accurate detection of the pupil diameter, as can be seen from the photos. Even for extreme orientations of the eye, the algorithm is able to capture the entire pupil in its depth. 
    Very minor to none adjustments were needed to be made in the threshold to capture all the pupils, which showed some promising results.

\begin{figure}[h]
  \centering
  \includegraphics[width=\linewidth]{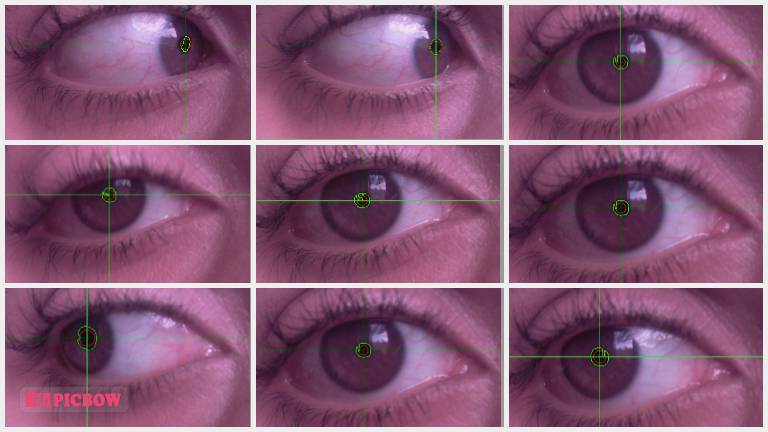}
  \caption{\textit{Arducam 5 MP Camera:} Images captured with the 5MP Arducam without using its IR source.}
  \label{fig:results_new_cam}
\end{figure}

    \item \textbf{Arducam with IR Source:} The performance of the algorithm is further boosted by the two IR sources used in this setup (see Figure \ref{fig:equipment_chart}. We can observe by just visual inspection that the images are more clear, enhanced and smooth with less noise. The IR lights are perfectly illuminating the eye and the pupil, which is then being detected through the voting mechanism of Hough's Transform. Similar to the last scenario, very minor adjustments in the threshold were made to capture these images.
    
\begin{figure}[h]
  \centering
  \includegraphics[width=\linewidth]{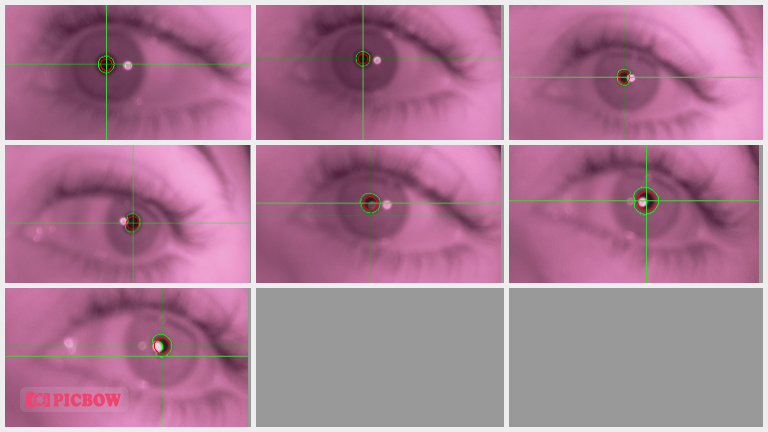}
  \caption{\textit{Arducam with IR Source:} Images captured with the 5MP Arducam using the two IR LED sources that come as a part its general assembly.}
  \label{fig:results_new_cam_night}
\end{figure}

     \item \textbf{Smart Phone:} Finally, we tried to run one round of experiments using the builtin camera module of Samsung S8. Contrary to our initial belief, the results obtained were slightly less than the Arducam's 5MP camera. Though the algorithm is able to capture most of the cases of pupil, it is very slightly off in two different scenarios. However, this should not be taken as a surprise because under normal camera settings, smart phones are not able to tap into additional light (contrary to long exposure modes); whereas the Arducam and Noir V2 cover this difference through tapping into the natural IR light present in the environment (having their IR filters removed). The IR filter in mobile phones do not allow for such experimentation leeway.
    
\begin{figure}[h]
  \centering
  \includegraphics[width=\linewidth]{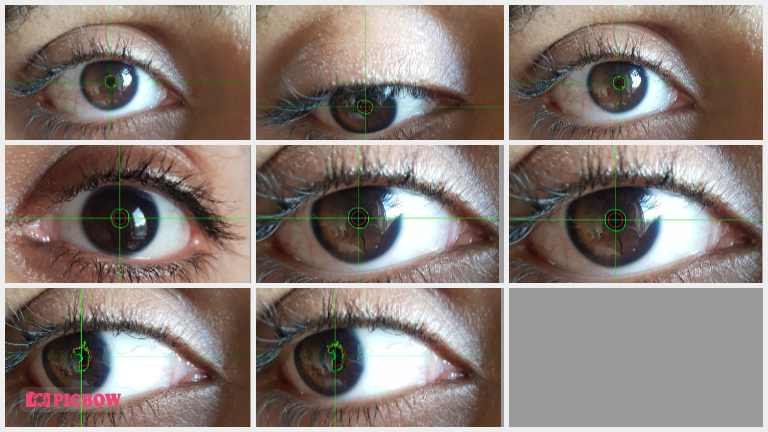}
  \caption{\textit{Smart Phone Camera:} Images captured through a Samsung S8 Smart Phone. Its inbuilt camera module is powered at 12MP with an F1.7 Aperture.}
  \label{fig:results_mobile}
\end{figure}

\end{enumerate}

Finally, we conclude this subsection by sharing the table for the diameter values of the captured images. Our simplistic and light-weight implementation of the algorithm yields good results and comparable accuracy to the pupillary size variations under a number of scenarios.

\begin{table*}[ht]
\centering
\begin{tabular}{||c|c|c|c|c|c|c|c||}
 \hline
 Experiment Setup & Image1 & Image2 & Image3 & Image4 & Image5 & Image6 & Image7 \\
 \hline\hline
 NOIR V2 Person One & - & - & - & - & - & - & - \\ 
 NOIR V2 Person Two & 57.10 & 111.50 & 112.76 & 129.33 & 158.10 & 166.50 & 184.49  \\
 NOIR V2 with IR Lamp & 127.82 & 173.88 & 174.90 & 189.69 & 206.96 & 220.50 & 224.60  \\
 Arducam 5MP Camera & 103.91 & 106.89 & 121.50 & 126.56 & 138.17 & 145.39 & 163.48 \\
 Arducam with IR Source & 92.36 & 95.17 & 95.96 & 110.20 & 137.22 & 138.34 & 142.66  \\
 Smart Phone Camera & 104.52 & 125.83 & 139.68 & 159.66 & 180.19 & 270.65 & 278.27  \\
 \hline
\end{tabular}
\caption{Table of pixelar diameter Values from images of each experimentation setup. Image 1 to 7 correspond to respective images in each of the results figure above, starting from top left in a chronological manner. This diameter was computed inversely by first calculating the contour area of each circle, and then the radius. Note. shaded pupil area of above may seem disproportionate with the values in this table due to cropping for images for this paper}
\label{table:results_diameter}
\end{table*}

\subsection{Discussion \& Concluding Remarks}
These experiments go on to show that the design parameters, defined earlier, pay a pivotal role in the successful operation of such a system. Though our initial work uses still images for post-processing, the same algorithmic pipeline can be adapted to videos. The overall detection capability of the system can be observed through the correlation of results with the diameter values. The cropping of images might create a disproportionate effect at first look, but a through run-through of the images displays that this modular mechanism is sensitive to slight variations in the eyes. And thus can be deployed for a number of applications, evident in the Figure \ref{fig:disease_chart}. 

There are a number of observations that can be inferred from all of these experiments. The lighting source plays a pivotal role in the process. Even though both the cameras were able to capture a significantly good quality of pictures, with one (Arducam) leading the other (Noir V2), however both the systems do not have enough resolute capabilities to capture the pupil in ultra-low to no light scenarios. Thus, an illumination source is required to meet the final target of a continuous pupil monitoring ecosystem. 

Additionally, the resolution of the camera doesn't hold much significance if there is not option of focal adjustment of the camera lens. Arducam was able to perform well in this regard owing to it's builtin feature of adjustable focus; thus leading to very clear and sharp images. Moreover, both the camera currently have an inch of a form factor. In order to deploy these cameras or similar to a pupil measuring task, the currently embedded circuit board needs to be moved further away from the camera head.

The overall system was currently being powered with a 5-volt USB-C cable tethered to a laptop. This can be easily replaced with an external battery source that provides a similar amount of voltage and current. Another option possible is to use the pupillography module in a regulated manner - i.e. ensuring sustained usage through interval of operations. As we can see from the Table \ref{table:results_diameter} as well, the resting diameter of the pupil has only slight variations. Thus, keeping this into account, the processing load of the system can be decreased further by only observing and storing the diameter of the pupils, and keep evaluating it in intervals with new values. 

The current approach of using Hough's transform for the said purpose is fairly light-weight. However, there are certain number of operations that can be further reduced to save computational power, and hence reduce energy requirements. In order to smooth out the images and reduce certain types of noise, we were implementing two filters in series. However, in the case of a auto-focus camera that creates smooth and sharp images, this requirement of filters might diminish. Hence, this should reduce the computation by at-least 10\% to 20\%. 

Lastly, the retinal color of the eye also impacts the detection efficiency of the system. Black colored eyes make a more challenging situation in general scenarios. But as mentioned earlier, a very small illumination source can overcome this design challenge. Moreover, the LED light source needs not to be too extensive. As we have shown in our implementation, even one or two small LEDs are more than enough for sufficiently lighting up the pupillay area. 
\section{Conclusion}

In this paper, we attempted to provide a brief background of the biology of our eyes and the current modes of measuring the pupillary reaction. Often undermined, these reactions can be the gateway to a wide range of early diagnostic applications which could potentially save thousands, if not millions, of lives. Here, we make the case of developing a visual ecosystem that revolves around continuous monitoring of eyes and pupils. With the wearable technology about to reach a market cap of USD 111.9 billion by 2028, its not too far away that such a technology would find its niche in the global market. We also present a number of design parameters that need to be accounted for, if aiming at building a compact, remote, cost effective and energy efficient system. Through our experimentation and results, we solidify these design constraints further, and share how different lighting conditions, retinal colors, resolution capacities, measurement distances, and eye orientation can lead to diverse set of challenges and varying results. However, we also prove through our algorithmic implementation that these can be overcome with a decent processing pipeline, and a good shortlisting of equipment. While our study observed many gaps between the actual design requirements and currently available off-the-shelf products, our findings propose a good way ahead in designing an ecosystem that revolves around visual health monitoring. We hope this paper helps to guide the directions for future research in the domains of pupillography and IoT, that benefit both, the end users and businesses involved.

\bibliographystyle{ACM-Reference-Format}
\bibliography{main_bib}

\end{document}